\documentclass[pra,aps,amsmath,amssymb,showkeys]{revtex4}
\usepackage{graphicx}

\newcommand{\pen}{\openone}
\newcommand{\msg}{{\mathsf{G}}}
\newcommand{\msh}{{\mathsf{H}}}
\newcommand{\msj}{{\mathsf{J}}}
\newcommand{\um}{{\mathsf{U}}}
\newcommand{\am}{{\mathsf{A}}}

\newcommand{\iu}{{\mathtt{i}}}
\newcommand{\cli}{{\mathcal{I}}}
\newcommand{\clm}{{\mathcal{M}}}

\newcommand{\cpt}{{\mathtt{C}}}
\newcommand{\wka}{\widetilde{k}}
\newcommand{\wel}{\widetilde{\ell}}

\newcommand{\bro}{\boldsymbol{\rho}}
\newcommand{\bdl}{\boldsymbol{\delta}}
\newcommand{\bsg}{\boldsymbol{\sigma}}
\newcommand{\vbro}{\boldsymbol{\varrho}}
\newcommand{\varep}{\varepsilon}
\newcommand{\tr}{\mathrm{tr}}
\newcommand{\dig}{\mathrm{diag}}

\newcommand{\suc}{\mathrm{suc}}
\newcommand{\rg}{\mathrm{g}}
\newcommand{\ron}{{\mathrm{ran}}}
\begin{document}
\clearpage
\preprint{}

\title{On the role of dealing with quantum coherence in amplitude amplification}

\author{Alexey E. Rastegin}
\affiliation{Department of Theoretical Physics, Irkutsk State University,
Gagarin Bv. 20, Irkutsk 664003, Russia}

\begin{abstract}
Amplitude amplification is one of primary tools in building
algorithms for quantum computers. This technique generalizes key
ideas of the Grover search algorithm. Potentially useful
modifications are connected with changing phases in the rotation
operations and replacing the intermediate Hadamard transform with
arbitrary unitary one. In addition, arbitrary initial distribution
of the amplitudes may be prepared. We examine trade-off relations
between measures of quantum coherence and the success probability
in amplitude amplification processes. As measures of coherence,
the geometric coherence and the relative entropy of coherence are
considered. In terms of the relative entropy of coherence,
complementarity relations with the success probability seem to be
the most expository. The general relations presented are
illustrated within several model scenarios of amplitude
amplification processes. 
\end{abstract}


\maketitle

\pagenumbering{arabic}
\setcounter{page}{1}

\section{Introduction}\label{sec1}

The Grover search algorithm \cite{grover97,grover97a,grover98} is
one of fundamental discoveries that motivate quantum
computations. Celebrated Shor's results \cite{shor97} have led to
numerous quantum algorithms for algebraic problems
\cite{hama2006,hallg07,vandam2010}. The authors of \cite{loka2007}
gave arguments that Grover's and Shor's algorithms are more
closely related than one might expect at first. It was soon
recognized that Grover's algorithm is optimal for searching by
queries to oracle \cite{bbbv97,zalka99}. Here, we invoke the
oracle to evaluate any item, whereas database {\it per se} is not
represented explicitly. Today, there exists a class of quantum
algorithms inspired by the Grover algorithm \cite{patel2016}.
Due to the broad applicability of search problems
\cite{patel2016}, researchers have attempted to formulate building
blocks of the algorithm as generally as possible. The original
algorithm may be modified by changing phases in the rotation
operations and replacing the intermediate Hadamard transform with
arbitrary unitary one. In addition, quantum computing may start with
an arbitrary initial distribution of the amplitudes. Details
of generalized versions of the Grover algorithm were
described in \cite{biham99,biham2000,biham2002}.

It is well known that entanglement is a key resource in quantum
information processing. The quantum parallelism of Deutsch
\cite{deutsch85} assumes to use entangled states of a quantum
register. The results of the papers \cite{bpati2002,jozsa03}
manifested that quantum speed-up without entanglement seems to be
impossible. On the other hand, quantum computations are connected
with only limited number of bases, in which states of the register
are represented. In order to analyze the quantum computational
speed-up, we should think about quantum correlations with respect
to the computational basis. The problem of quantifying coherence
at the quantum level is currently the subject of active researches
\cite{bcp14,abc16,plenio16,fan2017}. Studies of the role of quantum coherence in
performing quantum computations were reported in
\cite{hillery16,hfan2016,apati2016}. In particular, they ask
whether coherence may be a resource for increasing the power of
quantum algorithms. The authors of \cite{hfan2016} examined
coherence depletion in the original Grover algorithm. The paper
\cite{apati2016} is devoted to coherence and entanglement monogamy
in the discrete analogue of analog Grover search. Such analogues
of digital quantum computations were first addressed in
\cite{farhi98}.

The aim of this work is to study trade-off relations between
quantum coherence and the success probability in amplitude
amplification. The following aspects are mainly addressed. First,
more general scenarios of computing will be considered. Second, we
will formulate general relations between quantum coherence and the
success probability. Third, the geometric coherence will be
utilized in the algorithmic context. The paper is organized as
follows. In Sect. \ref{sec2}, we review the required material on
coherence quantifiers and techniques of amplitude amplification.
General relations between quantum coherence and the success
probability are considered in Sect. \ref{sec3}. In particular, we
derive a two-sided estimate on the relative entropy of coherence
in terms of the success probability. In Sect. \ref{sec4}, the
presented trade-off relations are exemplified  within some model
scenarios of amplitude amplification. In Sect. \ref{sec5}, we
conclude the paper with a summary of the results obtained.
Necessary results of solving recursion equations for a generalized
version of Grover's algorithm are listed in Appendix \ref{apa}.

\section{Preliminaries}\label{sec2}

In this section, we recall definitions of quantum coherence
quantifiers that will be used through the paper. We further
describe a generalized version of the Grover search algorithm.
Studying the role of quantum coherence in the context of amplitude
amplification, we typically refer to the computational basis. Let
the quantum register contain $n$ qubits. Then the basis is formed
by $N=2^{n}$ orthonormal kets. Each ket $|x\rangle$ is indexed by
binary $n$-string with $x_{j}\in\{0,1\}$. A rigorous framework for
the quantification of coherence was proposed in \cite{bcp14}. We
consider the set $\cli$ of all diagonal density matrices written
as
\begin{equation}
\bdl=\sum_{x=0}^{N-1} \delta_{x}\,|x\rangle\langle{x}|
\, . \label{incd}
\end{equation}
One further asks how far the given state is from states of the
form (\ref{incd}). The authors of \cite{bcp14} listed general
conditions for quantifiers of coherence. From a general
perspective, coherence measures are discussed in
\cite{abc16,plenio16,fan2017}. In the present paper, we will
mainly use the relative entropy of coherence and the geometric
coherence.

Using the quantum relative entropy as a measure of
distinguishability, we arrive at the relative entropy of
coherence. The quantum relative entropy of $\bro$ with respect
to $\bsg$ is defined as \cite{nielsen,watrous1}
\begin{equation}
D_{1}(\bro||\bsg):=
\begin{cases}
\tr(\bro\ln\bro-\bro\ln\bsg) \,,
& \text{if $\ron(\bro)\subseteq\ron(\bsg)$} \, , \\
+\infty\, , & \text{otherwise} \, .
\end{cases}
\label{relan}
\end{equation}
By $\ron(\bro)$, we mean here the range of $\bro$. The role of the
quantum relative entropy in quantum information science is
considered in \cite{nielsen,watrous1,vedral02}. The
relative-entropy-based measure of coherence is defined as
\cite{bcp14}
\begin{equation}
C_{1}(\bro):=
\underset{\bdl\in\cli}{\min}\,D_{1}(\bro||\bdl)
\, . \label{c1df}
\end{equation}
The minimization finally gives \cite{bcp14}
\begin{equation}
C_{1}(\bro)=S_{1}(\bro_{\dig})-S_{1}(\bro)
\, , \label{c1for}
\end{equation}
where $S_{1}(\bro)=\!{}-\tr(\bro\ln\bro)$ is the von Neumann
entropy, and $\bro_{\dig}$ is obtained from $\bro$ by vanishing
all off-diagonal elements calculated in the basis
$\bigl\{|x\rangle\bigr\}$. The first term in the right-hand side
of (\ref{c1for}) is equal to the Shannon entropy of the discrete
distribution with probabilities $p(x)=\langle{x}|\bro|x\rangle$,
viz.
\begin{equation}
S_{1}(\bro_{\dig})=H_{1}(p):=-\sum_{x=0}^{N-1} p(x)\,\ln{p}(x)
\, . \label{slin}
\end{equation}
The maximal value of (\ref{slin}) is equal to $\ln{N}$, so that
the quantity (\ref{c1for}) cannot exceed $\ln{N}$. Properties of
the relative entropy of coherence are discussed in
\cite{bcp14,plenio16,fan2017}. Uncertainty relations in terms of
the relative entropy of coherence were derived in
\cite{pati16,pzflf16,rastcomu}. Coherence monotones of the Tsallis
type were examined in \cite{rastpra16}. Relative R\'{e}nyi
entropies of coherence were considered in \cite{shao16,skwgb16}.
In contrast to (\ref{c1for}), these quantifiers do not reduce to a
simple analytical expression.

Several distance-based quantifiers of coherence were considered
\cite{bcp14,plenio16}. The $\ell_{1}$-norm of coherence is often
used as intuitively natural quantifier. For the given state
$\bro$, we define
\begin{equation}
C_{\ell_{1}}(\bro):=\underset{\bdl\in\cli}{\min}\,\bigl\|\bro-\bdl\bigr\|_{\ell_{1}}=
\sum_{x\neq{y}}\bigl|\langle{x}|\bro|y\rangle\bigr|
\, . \label{cel1}
\end{equation}
Complementarity and uncertainty relations in terms of the quantity
(\ref{cel1}) were studied in \cite{hall15,baietal6}. A
factorization relation for the $\ell_{1}$-norm of coherence was
proved in \cite{hufan15}. Another intuitively attractable way is
to use the squared $\ell_{2}$-norm. However, the corresponding
coherence quantifier does not satisfy the monotonicity requirement
\cite{bcp14}. The trace norm also induces an interesting candidate
to quantify the amount of coherence \cite{shao15,rpl15}.

We will use the geometric coherence which is introduced in terms
of the quantum fidelity \cite{uhlmann76,jozsa94}. Following Jozsa
\cite{jozsa94}, the fidelity of density matrices $\bro$ and $\bsg$
is expressed as
\begin{equation}
F(\bro,\bsg)=\bigl\|\sqrt{\bro}\,\sqrt{\bsg}\,\bigr\|_{1}^{2}
\, , \label{fiddf}
\end{equation}
where $\|\am\|_{1}=\tr\bigl(\sqrt{\am^{\dagger}\am}\,\bigr)$ is
the trace norm. Another known definition is written as the square
root of (\ref{fiddf}) \cite{nielsen,watrous1}. The fidelity ranges
between $0$ and $1$ taking the value $1$ for two identical states.
Using the unity minus fidelity as a distance measure, one defines
the geometric coherence by \cite{plenio16}
\begin{equation}
C_{\rg}(\bro):=1-\underset{\bdl\in\cli}{\max}\,F(\bro,\bdl)
\, . \label{cgdf}
\end{equation}
Properties of this coherence quantifier are summarized in
subsection III.C.3 of \cite{plenio16}. For pure states,
the formula (\ref{cgdf}) reduces to
\begin{equation}
C_{\rg}\bigl(|\psi\rangle\bigr)
=1-\underset{x}{\max}\,\bigl|\langle{x}|\psi\rangle\bigr|^{2}
\, . \label{cgpr}
\end{equation}
We will also use some bounds on the geometric coherence. To each
state $\bro$, we assign the index of coincidence
\begin{equation}
I(\bro):=\sum_{x=0}^{N-1} p(x)^{2}
\, , \label{icdf}
\end{equation}
where $p(x)=\langle{x}|\bro|x\rangle$. The authors of
\cite{geomes17} have proved that
\begin{equation}
\frac{N-1}{N}
\left\{1-
\sqrt{\,1-\frac{N}{N-1}\,\bigl[\tr(\bro^{2})-I(\bro)\bigr]}
\,\right\}
\leq{C}_{\rg}(\bro)
\, . \label{logeos}
\end{equation}
Using the unity minus the square root of (\ref{fiddf}), we also
have a distance measure. The corresponding coherence quantifier
was studied in \cite{bcp14,shao15}. Duality relations between the
coherence and path information were examined in
\cite{bera15,bagan16,qureshi17}. The authors of \cite{hufan16}
proposed the concept of relative quantum coherence.

Let us recall some results concerning generalizations of the Grover
algorithm. In our calculations, we will follow the scheme of
analysis developed in \cite{biham99,biham2000}. Suppose that the
search space contains $N=2^{n}$ items indexed by binary $n$-string
$x=(x_{1}\cdots{x}_{n})$ with $x_{j}\in\{0,1\}$ so that
$x\in\{0,1,\ldots,N-1\}$. The problem is to find one of marked
items which form the set $\clm$. In amplitude amplification, we
try to increase maximally amplitudes of states $|x\rangle$ just
for $x\in\clm$. Without loss of generality, the number of marked
items is assumed to obey $1\leq|\clm|\leq{N}/2$.

Each concrete item is tested by means of oracle-computable Boolean
function $x\mapsto{f}(x)$ such that $f(x)=1$ for $x\in\clm$ and
$f(x)=0$ for $x\in\clm^{\cpt}$. The original Grover algorithm
starts with initializing $n$-qubit register to $|0\rangle$ and
applying the Hadamard transform to get a uniform amplitude
distribution
\begin{equation}
\msh\,|0\rangle=\frac{1}{\sqrt{N}}\sum_{x=0}^{N-1} |x\rangle
\, . \label{uniam}
\end{equation}
Following \cite{biham99,biham2000}, we will use phase rotations
of two kinds. Rotations of the first kind
non-trivially act on unknown marked states. For an arbitrary phase
$\gamma$, we write
\begin{equation}
\msj_{f}(\gamma)=\sum_{x=0}^{N-1} \exp\bigl(\iu\gamma{f(x)}\bigr)\,|x\rangle\langle{x}|
\, . \label{rojf}
\end{equation}
Thus, the operation (\ref{rojf}) rotates all the marked states by
the phase $\gamma$. For $\gamma=\pi$, we have
$|x\rangle\mapsto-\,|x\rangle$ for $f(x)=1$, as in the original
formulation. Its realization with invoking the oracle black box is
well explained in the literature. Rotations of the second class
rotate some prescribed state $|s\rangle$ of the basis. For any
phase $\beta$, one reads
\begin{equation}
\msj_{s}(\beta)=\pen-\bigl(1-\exp(\iu\beta)\bigr)\,|s\rangle\langle{s}|
\, , \label{rojs}
\end{equation}
where $\pen$ is the identity operator. So, the ket $|s\rangle$ is
rotated by $\beta$, whereas other basis kets remain unchanged. One
of the well known ingredients of Grover's algorithm can be written
as
\begin{equation}
-\,\msj_{0}(\pi)=2\,|0\rangle\langle0|-\pen
\, . \label{roj0}
\end{equation}
Clothing (\ref{roj0}) by the Hadamard transforms, we obtain the
inversion about the average. So, the original Grover iteration is
written as
\begin{equation}
\msg_{0}=-\,\msh\,\msj_{0}(\pi)\,\msh\,\msj_{f}(\pi)
\, . \label{sgit0}
\end{equation}

For many reasons, generalizations of Grover's original
algorithm have been developed. In the generalized formulation, the
iteration (\ref{sgit0}) is replaced by
\begin{equation}
\msg=-\,\um\,\msj_{s}(\beta)\,\um^{\dagger}\,\msj_{f}(\gamma)
\, , \label{sgit1}
\end{equation}
with arbitrary $\beta$ and $\gamma$. Here, the Hadamard transform
on $n$ qubits is replaced by an arbitrary unitary operator. To the
given $|s\rangle$ and $\um$, we assign
$|\eta\rangle=\um\,|s\rangle$, whence
\begin{equation}
-\,\um\,\msj_{s}(\beta)\,\um^{\dagger}=\bigl(1-\exp(\iu\beta)\bigr)\,|\eta\rangle\langle\eta|-\pen
\, . \label{sgit2}
\end{equation}
In this way, the terms $\um$ and $|s\rangle$ are both taken into
account by $|\eta\rangle$ solely. After $t$ iterations, the state
of the quantum register is represented as
\begin{equation}
|g(t)\rangle=\sum_{x\in\clm} k_{x}(t)\,|x\rangle+\sum_{y\in\clm^{\cpt}} \ell_{y}(t)\,|y\rangle
\, . \label{gket0t}
\end{equation}
Iterations of the form (\ref{sgit1}) will be applied to an
arbitrary initial distribution of amplitudes $|g(0)\rangle$. It
must be stressed that initializing the state of the quantum
register can be quite challenging \cite{aasv2006}. Grover's search
algorithm for a mixed initial state of the register was analyzed
in \cite{biham2002}. Any single iteration changes amplitudes
according to the equations
\begin{equation}
k_{x}(t+1)=\langle{x}|\msg|g(t)\rangle
\, , \qquad
\ell_{y}(t+1)=\langle{y}|\msg|g(t)\rangle
\, , \nonumber
\end{equation}
where $x\in\clm$ and $y\in\clm^{\cpt}$. These equation are
accompanied by initial amplitudes $k_{x}(0)$ and $\ell_{y}(0)$ of
marked and unmarked items, respectively. The authors of
\cite{biham2000} have examined the above recursion equations.
Their results required for us are summarized in Appendix
\ref{apa}.

\section{General relations between quantum coherence and the success probability}\label{sec3}

In this section, we focus on general relations concerning
coherence changes in amplitude amplification. In practice,
computing devices are inevitably exposed to noise. In this
situation, the state of a quantum register after $t$ steps is
described by density matrix $\bro(t)$. Then the probability to
measure one of the marked state is written as
\begin{equation}
P_{\suc}(t)=\sum_{x\in\clm} \langle{x}|\bro(t)|x\rangle
\, . \label{renp01}
\end{equation}
Amplitude amplification processes aim to magnify amplitudes of
desired states unknown {\it a priori}. In order to increase the
success probability $P_{\suc}(t)$, a coherence of the register
should be used somehow. For original Grover's formulation, this
issue was addressed in \cite{hfan2016}. The writers of \cite{hfan2016} 
also concerned other quantum correlations such as pairwise or multipartite
entanglement and discord. We shall analyze a coherence behavior in more 
general setting. 

We begin with a simple situation, where the geometric coherence
immediately links to the success probability. For pure states, the
geometric coherence is expressed by (\ref{cgpr}). We first suppose
that only one amplitude should be maximized in a particular case.
This situation is sufficiently typical, but not general. For a
time, we also assume that states of a quantum register are pure
during the performance of algorithm. After a proper number of
iterations, amplitudes in some superposition $|\psi\rangle$ will
be sufficiently small, except for the unique one. We then have
\begin{equation}
C_{\rg}\bigl(|\psi\rangle\bigr)+P_{\suc}\bigl(|\psi\rangle\bigr)=1
\, . \label{cgrel}
\end{equation}
Under the described circumstances, an increase of the success
probability implies a decrease of the geometric coherence, and
{\it vice versa}. In the form of inequality, we can exceed the
above relation to mixed states of a quantum register and arbitrary
number of marked states.

\newtheorem{fip0}{Proposition}
\begin{fip0}\label{fip0lab}
Let $\bro$ be a density matrix normalized as $\tr(\bro)=1$, and
let $P_{\suc}(\bro)$ be defined according to (\ref{renp01}). The
geometric coherence of $\bro$ satisfies
\begin{equation}
\frac{N-1}{N}
\left\{1-
\sqrt{\,1-\frac{N}{N-1}\,\bigl[\tr(\bro^{2})-P_{\suc}^{2}-(1-P_{\suc})^{2}\bigr]}
\,\right\}\leq
C_{\rg}(\bro)\leq1-\frac{P_{\suc}}{M}
\ . \label{fidp01}
\end{equation}
\end{fip0}

{\bf Proof.} To prove the right-hand side of (\ref{fidp01}),
we consider the diagonal state
\begin{equation}
\bdl_{0}=\frac{1}{M}\sum_{x\in\clm} |x\rangle\langle{x}|
\, . \nonumber
\end{equation}
Using the property P4(b) of \cite{jozsa94}, we can write the
inequalities
\begin{equation}
\underset{\bdl\in\cli}{\max}\,F(\bdl,\bro)\geq
F(\bdl_{0},\bro)\geq\tr(\bdl_{0}\bro)=\frac{P_{\suc}}{M}
\ . \label{fimap}
\end{equation}
Combining (\ref{cgdf}) with (\ref{fimap}) completes the proof of
the upper bound.

Let us proceed to the left-hand side of (\ref{fidp01}). For all
$x\in\clm$ and $y\in\clm^{\cpt}$, one has $p(x)\leq{P}_{\suc}$ and
$p(y)\leq1-P_{\suc}$, so that
\begin{equation}
\sum_{x\in\clm}p(x)^{2}\leq{P}_{\suc}^{2}
\, , \qquad
\sum_{y\in\clm^{\cpt}}p(y)^{2}\leq(1-P_{\suc})^{2}
\, . \label{psmsmc}
\end{equation}
Hence, we obtain $I(\bro)\leq{P}_{\suc}^{2}+(1-P_{\suc})^{2}$.
Combining the latter with (\ref{logeos}) immediately gives the
left-hand side of (\ref{fidp01}). $\blacksquare$

The statement of Proposition \ref{fip0lab} is a two-sided estimate
of the geometric coherence in terms of the success probability
$P_{\suc}(\bro)$. During the proof, we have seen another relation
that deserves to be given explicitly. It follows from
(\ref{fimap}) that
\begin{equation}
P_{\suc}(\bro)\leq{M}\>\underset{\bdl\in\cli}{\max}\,F(\bdl,\bro)
\, . \label{cofimap}
\end{equation}
Operationally, this relation shows that the success probability is
bounded by the number of marked objects multiplied by the maximum
overlap of the state with an incoherent state. It should be noted
that the left-hand side of (\ref{fidp01}) provides a non-trivial
lower bound only when
\begin{equation}
P_{\suc}^{2}+(1-P_{\suc})^{2}\leq\tr(\bro^{2})
\, . \label{gecons}
\end{equation}
This condition always holds for pure states, since
$\tr(\bro^{2})=1$ for $\bro=|\psi\rangle\langle\psi|$. Returning
to the case of amplitude amplification, we see the following. If
the success probability is determined by the unique state of the
calculation basis, i.e., $M=1$, then
\begin{equation}
C_{\rg}(\bro)+P_{\suc}(\bro)\leq1
\, . \label{cgrelm}
\end{equation}
In reality, states of the quantum register are inevitably exposed
to noise. With some amount even small, they will become mixed. So,
the trade-off relation between quantum coherence and the success
probability may only be restricted here. Note that the geometric
coherence is not a coherence measure in the sense that is proposed
by the authors of \cite{plenio16}. They added the list of axioms
for coherence quantifiers by two items called the uniqueness for
pure states and the additivity. In this regard, we are also
interested in relations of the success probability with other
coherence quantifiers.

The coherence quantifier based on the relative entropy is one of
the most natural measure for these purposes. At the same time,
this measure is not connected with $P_{\suc}$ so immediately as
the geometric coherence. Here, the following statement holds.

\newtheorem{renp0}[fip0]{Proposition}
\begin{renp0}\label{renp0lab}
Let $\bro$ be a density matrix normalized as $\tr(\bro)=1$. For
the given $P_{\suc}(\bro)$ and $S_{1}(\bro)$, the relative entropy
of coherence satisfies the inequalities
\begin{align}
& h_{1}(P_{\suc})-S_{1}(\bro)\leq{C}_{1}(\bro)
\nonumber\\
&\leq
P_{\suc}\,\ln\!\left(\frac{M}{P_{\suc}}\right)+
(1-P_{\suc})\,\ln\!\left(\frac{N-M}{1-P_{\suc}}\right)-S_{1}(\bro)
\, , \label{renp02}
\end{align}
where $h_{1}(P_{\suc})$ is the binary Shannon entropy. Using the
argument of $\,\max{p}(x)=\max\langle{x}|\bro|x\rangle$, we
further define
\begin{equation}
\Omega:=
\begin{cases}
P_{\suc} \, ,
& \arg\bigl(\max{p}(x)\bigr)\in\clm \, , \\
1-P_{\suc}\, , & \text{\rm{otherwise}} \, .
\end{cases}
\label{omdef}
\end{equation}
Then the relative entropy of coherence satisfies
\begin{equation}
-\ln\Omega-S_{1}(\bro)\leq{C}_{1}(\bro)
\, . \label{renp03}
\end{equation}
\end{renp0}

{\bf Proof.} To prove the right-hand side of (\ref{renp02}), we
will use the results of \cite{popf2016}. The authors of
\cite{popf2016} addressed the situation, when independent events
are somehow collected into nonempty and pairwise disjoint subsets.
Upper bounds on the Shannon entropy of the original probability
distribution is then expressed in terms of new probabilities
corresponding to these subsets. In our case, we have the subsets
$\clm$ and $\clm^{\cpt}$ with probabilities $P_{\suc}$ and
$1-P_{\suc}$, respectively. Applying theorem 3 of \cite{popf2016}
to the probabilities $p(x)$, we write
\begin{equation}
H_{1}(p)\leq
P_{\suc}\,\ln\!\left(\frac{|\clm|}{P_{\suc}}\right)+
(1-P_{\suc})\,\ln\!\left(\frac{|\clm^{\cpt}|}{1-P_{\suc}}\right)
 . \nonumber
\end{equation}
Combining this with the definition of $C_{1}(\bro)$ completes the
proof of the right-hand side of (\ref{renp02}).

Let us proceed to the left-hand side of (\ref{renp02}). We first
observe that
\begin{equation}
\frac{1}{P_{\suc}}\sum_{x\in\clm}p(x)=1
\, , \qquad
\frac{1}{1-P_{\suc}}\sum_{y\in\clm^{\cpt}}p(y)=1
\, . \nonumber
\end{equation}
Applying Jensen's inequality to convex function
$\xi\mapsto-\ln\xi$ then gives
\begin{equation}
H_{1}(p)\geq
\!{}-P_{\suc}\,\ln\!\left(
\frac{1}{P_{\suc}}\sum_{x\in\clm}p(x)^{2}
\right)
-(1-P_{\suc})\,\ln\Biggl(
\frac{1}{1-P_{\suc}}\sum_{y\in\clm^{\cpt}}p(y)^{2}
\Biggr)
\, . \label{twos2}
\end{equation}
Combining (\ref{psmsmc}) with (\ref{twos2}) and decreasing of the
function $\xi\mapsto-\ln\xi$ finally gives
$H_{1}(p)\geq{h}_{1}(P_{\suc})$.

The proof of (\ref{renp03}) begins with applying Jensen's
inequality to convex function $\xi\mapsto-\ln\xi$, so that
\begin{equation}
-\sum_{x\in\clm}p(x)\,\ln{p}(x)
\geq
-\ln\!\left(
\sum_{x=0}^{N-1}p(x)^{2}
\right)
 . \label{twos23}
\end{equation}
If $\arg\bigl(\max{p}(x)\bigr)\in\clm$, then we have
$p(x)\leq\max{p}(x)\leq{P}_{\suc}$ for all
$x\in\{0,1,\ldots,N-1\}$, whence
\begin{equation}
\sum_{x=0}^{N-1}p(x)^{2}
\leq{P}_{\suc}
\, . \label{twos24}
\end{equation}
If $\arg\bigl(\max{p}(x)\bigr)\not\in\clm$, the right-hand side of
(\ref{twos24}) is replaced with $1-P_{\suc}$. Together with
(\ref{twos23}) and decreasing of $\xi\mapsto-\ln\xi$, these facts
provide (\ref{renp03}). $\blacksquare$

If we know $S_{1}(\bro)$, then Proposition
\ref{renp0lab} gives a two-sided estimate of the relative entropy
of coherence in terms of the success probability. Of course,
productive computations should keep states of the register closely
to the pure ones with small values of the von Neumann entropy.
When $P_{\suc}$ approaches $1$, a band of allowed values of
$C_{1}(\bro)$ becomes more and more narrow. This picture
characterizes good algorithms, which ensure high chances for the
success. In general, we see
some complementarity between coherence and the success
probability. For the fixed $P_{\suc}(\bro)$ and $S_{1}(\bro)$, a
difference between the lower and upper bounds depends on the ratio
$M/N$. The less this ratio is, the less width of coherence
variations is given by (\ref{renp02}).

When other terms are fixed, the right-hand side of (\ref{renp02})
decreases with $P_{\suc}$ in the interval $(M/N;1)$. The value
$M/N$ gives the success probability in the trivial algorithm of
random choice of items, when no amplification actually takes
place. Over this value, any increase of the success probability
will lead to decreasing range of allowed changes of quantum
coherence with respect to the computational basis. To reach
sufficiently high values of $P_{\suc}$, an algorithm of amplitude
amplification should somehow provide a coherence depletion. Note
also that the right-hand side of (\ref{renp02}) is saturated in
some examples of amplitude amplification. Thus, this upper bound
cannot be improved without considering additional parameters.

We also obtained two inequalities for estimating the relative
entropy of coherence from below. The formula (\ref{renp03}) can be
used, when we know in which of the sets $\clm$ and $\clm^{\cpt}$
the argument of $\max{p}(x)$ lies. This inequality may sometimes
give a stronger lower bound on the relative entropy of coherence
than (\ref{renp02}). In principle, the condition
$\arg\bigl(\max{p}(x)\bigr)\in\clm$ is sufficiently natural for a
good technique of amplitude amplification. We aim to build
algorithms that provide values $P_{\suc}>1/2$ and even as closely
to $1$ as possible. Since $M\leq{N}-M$, the above condition is
very plausible. Initially, this condition may be recorded {\it a
priori} in the initial amplitude distribution. When we do not have
data to apply the definition (\ref{omdef}), the left-hand side of
(\ref{renp02}) should solely be used as an estimate from below.

Among distance-based quantifiers, the $\ell_{1}$-norm of
coherence is one of most intuitive \cite{bcp14}. On the other
hand, it does not fulfill additional axioms proposed in
\cite{plenio16}. It seems that trade-off relations
between $C_{\ell_{1}}(\bro)$ and $P_{\suc}(\bro)$ are enough
complicated to formulate. Nevertheless, certain conclusions can be
obtained in some particular cases. In this regard, we recall the
corresponding discussion given in \cite{hfan2016}.

Suppose that amplitudes in the superposition
$|\psi\rangle=\sum_{x=0}^{N-1}c_{x}\,|x\rangle$ has the same
absolute values separately for labels in $\clm$ and $\clm^{\cpt}$,
namely
\begin{equation}
|c_{x}|=\alpha \quad \forall x\in\clm
\, , \qquad
|c_{y}|=\alpha^{\prime} \quad \forall y\in\clm^{\cpt}
\, . \label{cxcy}
\end{equation}
Such amplitude distribution is typical during a performance of the
standard Grover algorithm. In this case, we easily obtain
$\alpha^{2}=P_{\suc}/M$ and $(\alpha^{\prime})^{2}=(1-P_{\suc})/(N-M)$.
Combining the latter with the expression
\begin{equation}
1+C_{\ell_{1}}\!\bigl(|\psi\rangle\bigr)=\sum_{x,y=0}^{N-1} |c_{x}c_{y}|
\, , \nonumber
\end{equation}
we finally obtain
\begin{equation}
C_{\ell_{1}}\!\bigl(|\psi\rangle\bigr)=
\left(
\sqrt{MP_{\suc}}+\sqrt{(N-M)(1-P_{\suc})}
\,\right)^{2}-1
\, . \label{albe3}
\end{equation}
At first glance, a complementarity between the coherence
quantifier and the success probability is not obvious. Following
\cite{hfan2016}, we now take natural assumption $M\ll{N}$.
Then we rewrite (\ref{albe3}) in the form
\begin{equation}
C_{\ell_{1}}\!\bigl(|\psi\rangle\bigr)+NP_{\suc}=
N\bigl[1+O\bigl(\sqrt{M/N}\bigr)\bigr]
\, . \label{albe4}
\end{equation}
If $P_{\suc}$ is close to $1$ with $1-P_{\suc}=O(M/N)$, then the
right-hand side of (\ref{albe4}) can be replaced with
$N\bigl[1+O(M/N)\bigr]$. Rescaling by the denominator $N$, this
relation becomes very similar to (\ref{cgrel}) and (\ref{cgrelm}).
This rescaling could be expected here, since the $\ell_{1}$-norm
of coherence can increase up to $N-1$ \cite{hall15}, whereas the
geometric coherence cannot exceed $1$. In the context of original
Grover's algorithm, the relation (\ref{albe4}) was presented in
\cite{hfan2016}. We only note that the result (\ref{albe4})
reflects no more than a ``boxcar'' distribution of amplitudes.

In a similar manner, we could consider other situations of
amplitude amplification with {\it a priori} knowledge. In general,
however, complementarity relations between the $\ell_{1}$-norm of
coherence and the success probability are not easy to formulate.
Even the simplest non-trivial choice (\ref{cxcy}) has lead to
(\ref{albe3}). Despite of a simple structure of the density
matrix, the formula (\ref{albe3}) is complicated enough. Other
coherence quantifiers should be preferred, when we focus on
quantum coherence as a potential resource in amplitude
amplification. The statements of Propositions \ref{fip0lab} and
\ref{renp0lab} are formulated for an arbitrary quantum state. In
this sense, they may be not optimal for analyzing the change of
coherence just in amplitude amplification processes. Nevertheless,
we can use them whenever the actual density matrix or some of its
characteristics are known. We will present examples that
demonstrate how quantum coherence may be anti-correlated with the
success probability in amplitude amplification.

\section{Some model examples of coherence changes in amplitude amplifications}\label{sec4}

In this section, we will illustrate relations between coherence
and the success probability within explicit model examples. We
wish to study possible corollaries of the use of generalized
blocks in amplitude amplification from the viewpoint of their
bearing on quantum coherence. As we have recalled above, effects
of generalized blocks can be described by means of the single ket
$|\eta\rangle$. In the following, we will make some model
assumptions about $|\eta\rangle$ without a discussion of their
origin. We rather try to understand, whether quantum coherence is
a key resource in amplitude amplification. These studies allow us
to emphasize distinctions between cases, when marked and unmarked
states are dealt with consistently or inconsistently. For
convenience, we begin with the original Grover formulation
\cite{grover97}.

\subsection{Original Grover's formulation}\label{ssc41}

Let us set the initial amplitude distribution (\ref{uniam}) and
the rotation angles are $\beta=\gamma=\pi$. By $M=|\clm|$, we
mean the number of marked states, then $N-M=|\clm^{\cpt}|$. For
all $x\in\clm$ and $y\in\clm^{\cpt}$, the initial amplitudes
appear as
\begin{equation}
k_{x}(0)=\ell_{y}(0)=\frac{1}{\sqrt{N}}
\ . \label{init0}
\end{equation}
For the original algorithm, we have $|\eta\rangle=\msh\,|0\rangle$
and $\eta_{x}=\eta_{y}=1/\sqrt{N}$, whence
\begin{equation}
W_{k}=\frac{M}{N}
\ , \qquad
W_{\ell}=\frac{N-M}{N}
\ . \label{wtel0}
\end{equation}
Further, one gives $k_{x}^{\,\prime}(t)=\sqrt{N}\,k_{x}(t)$ and
$\ell_{y}^{\,\prime}(t)=\sqrt{N}\,\ell_{y}(t)$, including
$k_{x}^{\,\prime}(0)=\ell_{x}^{\,\prime}(0)=1$. In line with
(\ref{awka}) and (\ref{awel}), the initial weighted averages
appear as $\wka^{\,\prime}(0)=\wel^{\,\prime}(0)=1$, so that
\begin{equation}
\Delta{k}_{x}^{\,\prime}=\Delta\ell_{y}^{\,\prime}=0
\, . \label{ddel0}
\end{equation}
Then the formulas (\ref{delka1}) and (\ref{delel1}) merely say
that $k_{x}^{\,\prime}(t)=\wka_{x}^{\,\prime}(t)$ and
$\ell_{x}^{\,\prime}(t)=\wel_{x}^{\,\prime}(t)$ for all $t$. In
the original formulation, we actually deal only with two different
values of amplitudes.

Substituting $\beta=\gamma=\pi$ together with (\ref{wtel0}) into
(\ref{cosom}), we get
\begin{equation}
\cos\omega=W_{\ell}-W_{k}=\frac{N-2M}{N}
\ . \label{cosom0}
\end{equation}
Recall that we assume $1\leq{M}\leq{N}/2$. So, the parameter
$\omega$ is defined by (\ref{cosom}) and $\omega\in(0;\pi/2]$. It
will be convenient to remember the formulas
\begin{equation}
\sin^{2}\omega/2=\frac{M}{N}
\ , \qquad
\cos^{2}\omega/2=1-\frac{M}{N}
\ . \label{sicos0}
\end{equation}
For $\beta=\gamma=\pi$, we further have
$\omega_{\pm}=2\pi\pm\omega$ due to (\ref{eigen0}), whence the
eigenvalues (\ref{eigen0}) read as
\begin{equation}
\lambda_{\pm}=e^{\pm\iu\omega}
=\cos\omega\pm\iu\sin\omega
\, . \nonumber
\end{equation}
Further calculations lead to the following expressions,
\begin{equation}
\xi_{1}=\frac{-\iu\,\exp(+\iu\omega/2)}{2\sin\omega/2}
\ , \qquad
\xi_{2}=\frac{-\iu\,\exp(-\iu\omega/2)}{2\sin\omega/2}
\ .
\label{xi12fo}
\end{equation}
Using (\ref{cxi34}) together with $b=1+\cos\omega$ and
(\ref{xi12fo}), we obtain
\begin{equation}
\xi_{3}=\frac{\exp(+\iu\omega/2)}{2\cos\omega/2}
\ , \qquad
\xi_{4}=\frac{-\exp(-\iu\omega/2)}{2\cos\omega/2}
\ . \label{xi34fo}
\end{equation}
Since $t$ is integer, we may take $\omega_{\pm}=\!{}\pm\omega$
instead of $\omega_{\pm}=2\pi\pm\omega$. Due to (\ref{wkat}), one
has
\begin{equation}
\wka^{\,\prime}(t)=\frac{\sin\bigl[\omega(t+1/2)\bigr]}{\sin\omega/2}
\ , \qquad
\wel^{\,\prime}(t)=\frac{\cos\bigl[\omega(t+1/2)\bigr]}{\cos\omega/2}
\ . \label{wkwlfo}
\end{equation}
As was already mentioned, $k_{x}^{\,\prime}(t)=\wka^{\,\prime}(t)$
and $\ell_{y}^{\,\prime}(t)=\wel^{\,\prime}(t)$. Thus, we finally
write
\begin{equation}
k_{x}(t)=\frac{1}{\sqrt{N}}\>
\frac{\sin\bigl[\omega(t+1/2)\bigr]}{\sin\omega/2}
\ , \qquad
\ell_{y}(t)=\frac{1}{\sqrt{N}}\>
\frac{\cos\bigl[\omega(t+1/2)\bigr]}{\cos\omega/2}
\ , \nonumber
\end{equation}
for all $x\in\clm$ and $y\in\clm^{\cpt}$. The probabilities of
interest appear as
\begin{align}
P_{\suc}(t)&=\sum_{x\in\clm}|k_{x}(t)|^{2}=
\frac{M}{N\sin^{2}\omega/2}\,\sin^{2}\bigl[\omega(t+1/2)\bigr]
=\sin^{2}\bigl[\omega(t+1/2)\bigr]
\, , \label{psuc00}\\
1-P_{\suc}(t)&=\sum_{y\in\clm^{\cpt}}|\ell_{y}(t)|^{2}=
\frac{N-M}{N\cos^{2}\omega/2}\,\cos^{2}\bigl[\omega(t+1/2)\bigr]
=\cos^{2}\bigl[\omega(t+1/2)\bigr]
\, . \label{npsuc00}
\end{align}
We wish to relate these probabilities with the coherence
quantifiers. Since the state of the register is pure, its von
Neumann entropy is zero. Further, the diagonal part of the density
matrix reads as
\begin{equation}
\bro_{\dig}=\dig
\bigl(\,
\overset{\text{$M$ entries}}{\overbrace{|k_{x}(t)|^{2},\ldots,|k_{x}(t)|^{2}}},
\,\underset{\text{$N-M$ entries}}{\underbrace{|\ell_{y}(t)|^{2},\ldots,|\ell_{y}(t)|^{2}}}
\,\bigr)
\, . \label{dig00}
\end{equation}
Hence, the geometric coherence obeys the complementarity relation
\begin{equation}
C_{\rg}\bigl(|g(t)\rangle\bigr)=1-\frac{P_{\suc}(t)}{M}
\ , \label{fidp01fo}
\end{equation}
whenever $|k_{x}(t)|^{2}\geq|\ell_{y}(t)|^{2}$. Otherwise, the
geometric coherence is equal to the $1$ minus
$\bigl(1-P_{\suc}(t)\bigr)/(N-M)$. Further, the relative entropy
of coherence is calculated as
\begin{equation}
C_{1}\bigl(|g(t)\rangle\bigr)=\!{}-M|k_{x}(t)|^{2}\ln\!\left(|k_{x}(t)|^{2}\right)
-(N-M)|\ell_{y}(t)|^{2}\ln\!\left(|\ell_{y}(t)|^{2}\right)
 . \label{coh00}
\end{equation}
Due to (\ref{sicos0}), we can write
\begin{align}
M|k_{x}(t)|^{2}&=\sin^{2}\bigl[\omega(t+1/2)\bigr]
=P_{\suc}(t)
\, , \label{mknml1}\\
(N-M)|\ell_{y}(t)|^{2}&=\cos^{2}\bigl[\omega(t+1/2)\bigr]
=1-P_{\suc}(t)
\, . \label{mknml2}
\end{align}
Combining (\ref{coh00}) with the last two formulas gives
\begin{equation}
C_{1}\bigl(|g(t)\rangle\bigr)=
\!{}-P_{\suc}(t)\,\ln\!\left(\frac{P_{\suc}(t)}{M}\right)
-\bigl(1-P_{\suc}(t)\bigr)\ln\!\left(\frac{1-P_{\suc}(t)}{N-M}\right)
 . \label{coh000}
\end{equation}
The latter coincides with the right-hand side of
(\ref{renp02}). In contrast to (\ref{fidp01fo}), the formula
(\ref{coh000}) holds for all $t$. Therefore, the upper bound given
by (\ref{renp02}) is saturated in the original formulation of
Grover's search. For the given $P_{\suc}$, the relative entropy of
coherence reaches the maximal value approved by the right-hand
side of (\ref{renp02}). Grover's search algorithm works so that
any coherence decreasing is used in the most efficient way.

\subsection{Marked and unmarked states are consistently amplified or decayed}\label{ssc42}

Let us consider the case, when the terms $\um$ and $|s\rangle$ are
such that marked states are amplified or decayed consistently. In
this particular situation, some prior knowledge is available to
users. We assume that amplitudes of the state $|\eta\rangle$ read
as
\begin{equation}
\eta_{x}=\sqrt{\frac{M_{\eta}}{MN}}
\ , \qquad
\eta_{y}=\sqrt{\frac{N-M_{\eta}}{(N-M)N}}
\ , \label{etxy}
\end{equation}
where $x\in\clm$ and $y\in\clm^{\cpt}$. The angles $\beta$ and
$\gamma$ are the same as in the original formulation. When
amplitudes are balanced with respect to both $\clm$ and
$\clm^{\cpt}$, we describe them by a single non-integer parameter
$M_{\eta}>0$. An efficiency is increased with $M_{\eta}>M$ and
decreased with $M_{\eta}<M$. Here, we replace (\ref{wtel0}) with
the formulas
\begin{equation}
W_{k}=\frac{M_{\eta}}{N}
\ , \qquad
W_{\ell}=\frac{N-M_{\eta}}{N}
\ . \label{wtelb}
\end{equation}
Combining (\ref{etxy}) with (\ref{awka}) and (\ref{awel}) directly
gives $k_{x}^{\,\prime}(t)=\wka^{\,\prime}(t)$ for $x\in\clm$ and
$\ell_{x}^{\,\prime}(t)=\wel^{\,\prime}(t)$ for $y\in\clm^{\cpt}$.
Rewriting (\ref{cosom}) with $M_{\eta}$ instead of $M$, we have
the parameter $\omega_{\eta}$ such that
\begin{equation}
\sin^{2}\omega_{\eta}/2=\frac{M_{\eta}}{N}
\ , \qquad
\cos^{2}\omega_{\eta}/2=1-\frac{M_{\eta}}{N}
\ . \label{sicosb}
\end{equation}
The latter should be used simultaneously with (\ref{sicos0}). The
initial amplitude distribution (\ref{init0}) is the same as in the
original formulation. Correspondingly to (\ref{etxy}), we have
\begin{equation}
k_{x}^{\,\prime}(t)=\sqrt{\frac{MN}{M_{\eta}}}\> k_{x}(t)
\, , \qquad
\ell_{y}^{\,\prime}(t)=\sqrt{\frac{(N-M)N}{N-M_{\eta}}}\> \ell_{y}(t)
\, . \label{klrestb}
\end{equation}
Together with $k_{x}(0)=\ell_{y}(0)=1/\sqrt{N}$, one then obtains
\begin{align}
k_{x}^{\,\prime}(0)&=\wka^{\,\prime}(0)=\sqrt{\frac{M}{M_{\eta}}}
=\frac{\sin\omega/2}{\sin\omega_{\eta}/2}
\ , \label{wka0et}\\
\ell_{x}^{\,\prime}(0)&=\wel^{\,\prime}(0)=\sqrt{\frac{N-M}{N-M_{\eta}}}
=\frac{\cos\omega/2}{\cos\omega_{\eta}/2}
\ , \label{wel0et}
\end{align}
where the parameter $\omega$ is again defined by (\ref{cosom0}).

For $\beta=\gamma=\pi$, we further have
$\omega_{\pm}=2\pi\pm\omega_{\eta}$. As calculations show, the
formulas (\ref{xi12fo}) and (\ref{xi34fo}) are replaced with
\begin{align}
\xi_{1}&=\frac{-\iu\,\exp(+\iu\omega/2)}{2\sin\omega_{\eta}/2}
\, , &
\xi_{2}&=\frac{-\iu\,\exp(-\iu\omega/2)}{2\sin\omega_{\eta}/2}
\, ,
\label{xi12et}\\
\xi_{3}&=\frac{\exp(+\iu\omega/2)}{2\cos\omega_{\eta}/2}
\ , &
\xi_{4}&=\frac{-\exp(-\iu\omega/2)}{2\cos\omega_{\eta}/2}
\ . \label{xi34et}
\end{align}
Similarly to (\ref{wkwlfo}), we obtain the averaged amplitudes
\begin{equation}
\wka^{\,\prime}(t)=
\frac{\sin\bigl(\omega_{\eta}t+\omega/2\bigr)}{\sin\omega_{\eta}/2}
\ , \qquad
\wel^{\,\prime}(t)=
\frac{\cos\bigl(\omega_{\eta}t+\omega/2\bigr)}{\cos\omega_{\eta}/2}
\ . \label{wkwket}
\end{equation}
These formulas also represent the amplitudes
$k_{x}^{\,\prime}(t)=\wka^{\,\prime}(t)$ for $x\in\clm$ and
$\ell_{x}^{\,\prime}(t)=\wel^{\,\prime}(t)$ for $y\in\clm^{\cpt}$.
Substituting $t=0$, they reduce to (\ref{wka0et}) and
(\ref{wel0et}). Combining (\ref{klrestb}) with (\ref{wkwket})
immediately gives
\begin{equation}
|k_{x}(t)|^{2}=\frac{\sin^{2}\bigl(\omega_{\eta}t+\omega/2\bigr)}{M}
\ , \qquad
|\ell_{y}(t)|^{2}=\frac{\cos^{2}\bigl(\omega_{\eta}t+\omega/2\bigr)}{N-M}
\ . \nonumber
\end{equation}
Calculating the probabilities then results in
\begin{equation}
P_{\suc}(t)=\sin^{2}\bigl(\omega_{\eta}t+\omega/2\bigr)
\, , \qquad
1-P_{\suc}(t)=\cos^{2}\bigl(\omega_{\eta}t+\omega/2\bigr)
\, . \label{twopb}
\end{equation}
When $|k_{x}(t)|^{2}\geq|\ell_{y}(t)|^{2}$, the geometric
coherence again obeys (\ref{fidp01fo}). We also note that
relations (\ref{mknml1}) and (\ref{mknml2}) are still valid. The
relative entropy of coherence is again connected with the
probabilities by (\ref{coh000}). The success probability is first
maximized, when $\omega_{\eta}t+\omega/2$ becomes as close to
$\pi/2$ as possible. As $t$ is integer, we take one of the two
numbers
\begin{equation}
\left\lfloor
\frac{\pi-\omega}{2\,\omega_{\eta}}
\right\rfloor
 , \qquad
\left\lceil
\frac{\pi-\omega}{2\,\omega_{\eta}}
\right\rceil
 . \label{twoin}
\end{equation}
With growth of $M_{\eta}>M$ in (\ref{etxy}), the key parameter
$\omega_{\eta}>\omega$ also increases. It is natural that
both $M$ and $M_{\eta}$ are very small in comparison with $N$.
Then we approximately write
$\omega_{\eta}\approx2\sqrt{M_{\eta}/N}$. Hence, the optimal
measurement time can be estimated. When $M_{\eta}$ grows and other
parameters are fixed, the integers (\ref{twoin}) decrease
proportionally to $M_{\eta}^{-1/2}$. By $\omega_{\eta}$, one
characterizes a rate of amplitude amplification. Increasing
$M_{\eta}$ due to conducive prior knowledge, one reduces the
number of iterations required for maximizing the success
probability. For the fixed $M$ and $N$, we herewith accelerate an
algorithm work. When $M_{\eta}<M$, the situation is opposite. As
prior knowledge now prevents, reaching the maximum of the success
probability will demand more iterations. In both these situations,
the quantity $C_{1}\bigl(|g(t)\rangle\bigr)$ decreases with
increasing $P_{\suc}(t)$ according to the right-hand side of
(\ref{renp02}). Similarly to the original formulation, any
coherence reducing is used most efficiently. This feature holds
due to the consistency during the computing process. It evolves so
that states are evenly amplified in $\clm$ and attenuated in
$\clm^{\cpt}$.

Similar observations can be made, when the initial
amplitude distribution contains consistent prior knowledge.
The authors of \cite{hfan2016} addressed the case of arbitrary 
initial amplitudes in the original version. Together with coherence
depletion, a behavior of the optimal measurement time was studied. 
We only mention a few aspects that
were not addressed therein. Let $k_{x}(0)$ be independent of
$x\in\clm$ and $\ell_{y}(0)$ be independent of $y\in\clm^{\cpt}$,
but $k_{x}(0)\neq\ell_{y}(0)$. We can see that an amplification
process is again balanced so that the right-hand side of
(\ref{renp02}) is saturated. This property takes place, even if
$k_{x}(0)<\ell_{y}(0)$. This case also reveals the role of
consistency in amplitude amplification. Indeed, states will evenly
be amplified in $\clm$ and attenuated in $\clm^{\cpt}$. We refrain
from presenting the details here.

\subsection{Marked and unmarked states are inconsistently amplified or decayed}\label{ssc43}

We now concern the case, when consistency of states during
amplitude amplification is broken. Suppose that the amplitudes of
$|\eta\rangle$ can take only two different values written as
\begin{equation}
\sqrt{\frac{1+\alpha}{N}}
\ , \qquad
\sqrt{\frac{1-\alpha}{N}}
\ , \label{varpm}
\end{equation}
where $\alpha\in[0;1)$. Taking even $M$, we assume amplitudes to
be distributed as follows. For $x\in\clm$, there are $M/2$ values
$\eta_{x}=\sqrt{(1+\alpha)/N}$ and $M/2$ values
$\eta_{x}=\sqrt{(1-\alpha)/N}$. In effect, we also put
$\eta_{y}=\sqrt{(1+\alpha)/N}$ for one half and
$\eta_{y}=\sqrt{(1-\alpha)/N}$ for other half of items
$y\in\clm^{\cpt}$. Then the weights $W_{k}$ and $W_{\ell}$ satisfy
(\ref{wtel0}).

\begin{figure}
\includegraphics[width=8.0cm]{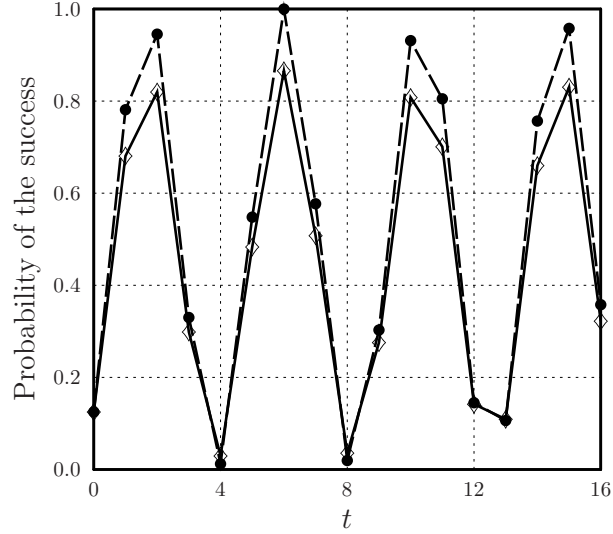}
\caption{\label{fig1} $P_{\suc}(t)$ as a function of integer $t$
is shown by solid line for $\alpha=0.72$ and by dashed line for
$\alpha=0$. Both the lines are related to the case $N=16$ and
$M=2$.}
\end{figure}

For the uniform initial distribution, the rescaled initial
amplitudes are written as
\begin{equation}
k_{x}^{\,\prime}(0)=\frac{1}{\sqrt{1\pm\alpha}}
\ , \qquad
\ell_{y}^{\,\prime}(0)=\frac{1}{\sqrt{1\pm\alpha}}
\ . \nonumber
\end{equation}
By substituting, the initial averaged amplitudes then read as
\begin{equation}
\wka^{\,\prime}(0)=
\wel^{\,\prime}(0)=
\frac{\sqrt{1+\alpha}+\sqrt{1-\alpha}}{2}
\ . \nonumber
\end{equation}
Unlike the above balanced cases, the initial differences take
non-zero values, namely
\begin{equation}
\Delta{k}_{x}^{\,\prime}=\Delta\ell_{y}^{\,\prime}=
\frac{1}{\sqrt{1\pm\alpha}}-\frac{\sqrt{1+\alpha}+\sqrt{1-\alpha}}{2}
\ . \label{delnz}
\end{equation}
Since $\wka^{\,\prime}(0)=\wel^{\,\prime}(0)$, we can obtain the
coefficients $\xi_{1}$, $\xi_{2}$, $\xi_{3}$, $\xi_{4}$ as
follows. The expressions (\ref{xi12fo}) and (\ref{xi34fo}) should
all be multiplied by the factor
$\bigl(\sqrt{1+\alpha}+\sqrt{1-\alpha}\,\bigr)/\sqrt{2}$. Hence,
the averaged amplitudes become
\begin{align}
\wka^{\,\prime}(t)&=\frac{\sqrt{1+\alpha}+\sqrt{1-\alpha}}{2}
\ \frac{\sin\bigl[\omega(t+1/2)\bigr]}{\sin\omega/2}
\ , \nonumber\\
\wel^{\,\prime}(t)&=\frac{\sqrt{1+\alpha}+\sqrt{1-\alpha}}{2}
\ \frac{\cos\bigl[\omega(t+1/2)\bigr]}{\cos\omega/2}
\ . \nonumber
\end{align}
Hence, we obtain the amplitudes of marked and unmarked states in
the form
\begin{align}
k_{x}(t)&=\sqrt{\frac{1\pm\alpha}{N}}
\left(
\wka^{\,\prime}(t)+\frac{1}{\sqrt{1\pm\alpha}}-
\frac{\sqrt{1+\alpha}+\sqrt{1-\alpha}}{2}
\,\right)
 , \label{kxtun}\\
\ell_{y}(t)&=\sqrt{\frac{1\pm\alpha}{N}}
\left(
\wel^{\,\prime}(t)+\frac{(-1)^{t}}{\sqrt{1\pm\alpha}}-
\frac{\sqrt{1+\alpha}+\sqrt{1-\alpha}}{2\,(-1)^{t}}
\,\right)
 . \label{lytun}
\end{align}
For each of the lines (\ref{kxtun}) and (\ref{lytun}), we have the
sign plus and the sign minus exactly for one half of the
amplitudes. By calculations, the desired probabilities are finally
expressed as follows. For $\varep=\pm$, we put the quantities
\begin{align}
P^{(\varep)}(t)
&=\frac{1}{2}
\left[
\,\sin\frac{\omega}{2}+
\bigl(1+\varep\alpha+\sqrt{1-\alpha^{2}}\,\bigr)
\sin\frac{\omega{t}}{2}\,\cos\frac{\omega(t+1)}{2}
\,\right]^{2}
 , \label{psunds}\\
Q^{(\varep)}(t)
&=\frac{1}{2}
\left[
\,\cos\frac{\omega}{2}-
\bigl(1+\varep\alpha+\sqrt{1-\alpha^{2}}\,\bigr)
\sin\frac{\omega{t}}{2}\,\sin\frac{\omega(t+1)}{2}
\,\right]^{2}
 , \label{qsunde}\\
Q^{(\varep)}(t)
&=\frac{1}{2}
\left[
\,\cos\frac{\omega}{2}-
\bigl(1+\varep\alpha+\sqrt{1-\alpha^{2}}\,\bigr)
\cos\frac{\omega{t}}{2}\,\cos\frac{\omega(t+1)}{2}
\,\right]^{2}
 , \label{qsundo}
\end{align}
where (\ref{qsunde}) stands for even $t$ and (\ref{qsundo}) stands
for odd $t$. In terms of these quantities, the probabilities of
interest are finally expressed as
\begin{equation}
P_{\suc}(t)=P^{(+)}(t)+P^{(-)}(t)
\, , \qquad
1-P_{\suc}(t)=Q^{(+)}(t)+Q^{(-)}(t)
\, . \label{mipun}
\end{equation}
An algorithm rate is characterized by the two parameters $\omega$
and $\alpha$. The former corresponds to the original version,
whereas the latter reflects the role of prior knowledge. Further,
the relative entropy of coherence is represented as
\begin{equation}
C_{1}\bigl(|g(t)\rangle\bigr)=\sum_{\varep=\pm}
P^{(\varep)}(t)\,\ln\biggl(\frac{M/2}{P^{(\varep)}(t)}\biggr)
+\sum_{\varep=\pm}
Q^{(\varep)}(t)\,\ln\biggl(\frac{(N-M)/2}{Q^{(\varep)}(t)}\biggr)
\, . \label{recun}
\end{equation}
Since the action of (\ref{sgit2}) is not balanced, the relation
between $C_{1}\bigl(|g(t)\rangle\bigr)$ and $P_{\suc}(t)$ does not
follow (\ref{coh000}). Due to an inconsistency in amplitude
amplification, coherence reducing is not used in the most
efficient way. A similar picture occurs with inconsistent prior
knowledge in the initial distribution. We refrain from presenting
the details here.

To illustrate the above conclusions, we now visualize
$C_{1}\bigl(|g(t)\rangle\bigr)$ versus $t$ for a some simple
choice of parameters. We also present both the bounds of the
two-sided estimation
\begin{align}
&\max\bigl\{h_{1}\bigl(P_{\suc}(t)\bigr),-\ln\Omega(t)\bigr\}\leq{C}_{1}\bigl(|g(t)\rangle\bigr)
\, , \label{twosit1}\\
&C_{1}\bigl(|g(t)\rangle\bigr)\leq
P_{\suc}(t)\,\ln\!\left(\frac{M}{P_{\suc}(t)}\right)
+\bigl(1-P_{\suc}(t)\bigr)\ln\!\left(\frac{N-M}{1-P_{\suc}(t)}\right)
 . \label{twosit2}
\end{align}
With changes of $t$, the condition
$\arg\bigl(\max{p}(x)\bigr)\in\clm$ is violated from time to time.

\begin{figure}
\includegraphics[width=8.0cm]{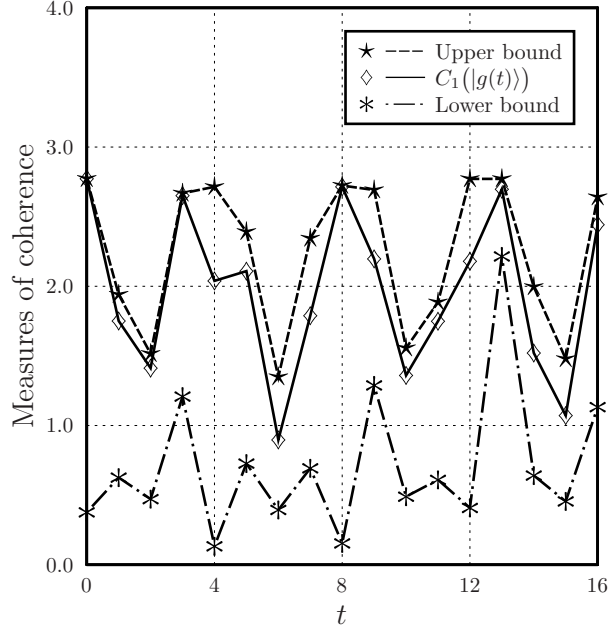}
\caption{\label{fig2} The relative entropy of coherence
$C_{1}\bigl(|g(t)\rangle\bigr)$ and bounds on it as functions of
$t$ for the case $N=16$, $M=2$, and $\alpha=0.72$.}
\end{figure}

In Figs. \ref{fig1} and \ref{fig2}, we visualize the results for
the case $N=16$, $M=2$, and $\alpha=0.72$. In Fig. \ref{fig1}, the
success probability is shown as a function of $t$. For comparison,
the success probability is also given for $\alpha=0$. It seems
that small variations of $\alpha$ do not alter essentially the
optimal measurement time. We also observe that an inconsistency of
amplitudes of the state $|\eta\rangle$ leads to some decreasing of
the probability peaks. On the other hand, reaching one or another
peak requires the same number of iteration as in the original
formulation. We also observe that the trade-off line between
$C_{1}\bigl(|g(t)\rangle\bigr)$ and the success probability does
not follow the upper bound exactly. Dealing with marked and
unmarked states inconsistently, this form of amplitude
amplification cannot always use coherence changes in the most
efficient way in the sense of impact on the success probability.
Using Fig. \ref{fig2}, we can also estimate a quality of bounds in
the two-sided estimate (\ref{renp02}). As explicit general bounds,
they seem to be sufficiently useful.

\subsection{An example of mixed states in original Grover's formulation}\label{ssc414}

In this section, we briefly exemplified the statements of
Propositions \ref{fip0lab} and \ref{renp0lab} in application to
mixed states. We restrict a consideration to impure states of the
form
\begin{equation}
\vbro(\vartheta)=(1-\vartheta)|\nu\rangle\langle\nu|+\vartheta|\mu\rangle\langle\mu|
\, ,\label{vmunu}
\end{equation}
where $\vartheta\in[0;1]$ and the normalized pure states
$|\nu\rangle$ and $|\mu\rangle$ are defined as
\begin{equation}
|\nu\rangle:=\frac{1}{\sqrt{N-M}}\sum_{y\in\clm^{\cpt}} |y\rangle
\, , \qquad
|\mu\rangle:=\frac{1}{\sqrt{M}}\sum_{x\in\clm} |x\rangle
\, . \label{smar}
\end{equation}
For such states, we can express results analytically. First of
all, we observe the equality
$\msg_{0}\,\vbro(\vartheta)\,\msg_{0}=\vbro(\vartheta)$. In other
words, states of the form (\ref{vmunu}) are not changed during the
Grover search. It is also obvious that $P_{\suc}(t)=\vartheta$.
Doing some calculations, we further obtain the result
\begin{equation}
\underset{\bdl\in\cli}{\max}\,
\bigl\|\sqrt{\vbro(\vartheta)}\,\sqrt{\bdl}\,\bigr\|_{1}=\sqrt{\frac{\vartheta}{M}+\frac{1-\vartheta}{N-M}}
\ , \qquad
C_{\rg}\bigl(\vbro(\vartheta)\bigr)=1-\frac{\vartheta}{M}-\frac{1-\vartheta}{N-M}
\ . \label{vbgc}
\end{equation}
Comparing the latter with (\ref{fidp01}), we see the following.
The left-hand side of (\ref{fidp01}) vanishes, whereas the
right-hand side excesses $C_{\rg}\bigl(\vbro(\vartheta)\bigr)$ by
the term $(1-\vartheta)\big/(N-M)$. It is a typical case, when the number
of marked states is very small in comparison with the total search
space. In this situation, the upper bound of Proposition
\ref{fip0lab} becomes almost tight. It is also obvious that
$S_{1}\bigl(\vbro(\vartheta)\bigr)=h_{1}(P_{\suc})$. Thus, the
left-hand side of the relation (\ref{renp02}) vanishes with states of
the form (\ref{vmunu}). Although we see some anti-correlation
between the coherence measures and the success probability with
respect to varying $\vartheta$, for each concrete
$\vbro(\vartheta)$ the success probability is constant. The latter
may change during amplitude amplification with generalized blocks
that deal with marked and unmarked states consistently. Thus,
quantum coherence alone is insufficient as a resource for the
original Grover algorithm.

\section{Conclusions}\label{sec5}

We have addressed the role of dealing with quantum coherence in
amplitude amplification processes. General trade-off relations
between quantum coherence and the success probability were
derived. It seems that the geometric coherence and the relative
entropy of coherence are more convenient quantifiers in this
context. In comparison with the geometric coherence, the relative
entropy of coherence is recognized as more sensitive. We obtained
inequalities that can be used for estimating the relative entropy
of coherence. Basic conclusions are supported by explicit
consideration of several model scenarios of amplitude
amplification. Coherence changes can be used in the most efficient
way only when marked and unmarked states are dealt with
consistently. In other words, the computing process does not
distinguish between the marked states as well as between the
unmarked ones. Then any coherence decreasing will amplify the
success probability as much as possible. Otherwise, there is a
difference between concrete marked states and, maybe,
unmarked states. In this unbalanced case, the relative
entropy of coherence does not reach always its maximal value
approved by the two-sided estimate for the given value of the
success probability. Our results evidence that even tight
trade-offs between coherence and the success probability do not
imply always an enhancement of amplitude amplification.

\appendix

\section{Results of analyzing the recursion equations}\label{apa}

In this section, we briefly recall some formulas for the
amplitudes in a generalized version of Grover's search with
arbitrary initial distribution \cite{biham2000}. The solution is
expressed in terms of rescaled amplitudes
\begin{equation}
k_{x}^{\,\prime}(t)=\eta_{x}^{-1}k_{x}(t)
\, , \qquad
\ell_{y}^{\,\prime}(t)=\eta_{y}^{-1}\ell_{y}(t)
\, . \label{resca}
\end{equation}
where the coefficients $\eta_{z}=\langle{z}|\eta\rangle$ are
assumed to be nonzero for all $z\in\clm\cup\clm^{\cpt}$.
Introducing the weights
\begin{equation}
W_{k}=\sum_{x\in\clm}|\eta_{x}|^{2}
\, , \qquad
W_{\ell}=\sum_{y\in\clm^{\cpt}}|\eta_{y}|^{2}
\, , \label{wkael}
\end{equation}
so that $W_{k}+W_{\ell}=\langle\eta|\eta\rangle=1$, the authors of
\cite{biham2000} defined weighted averages of rescaled amplitudes
\begin{align}
\wka^{\,\prime}(t)&=
\frac{1}{W_{k}}\sum_{x\in\clm}|\eta_{x}|^{2}\,k_{x}^{\,\prime}(t)
\, , \label{awka}\\
\wel^{\,\prime}(t)&=
\frac{1}{W_{\ell}}\sum_{y\in\clm^{\cpt}}|\eta_{y}|^{2}\,\ell_{y}^{\,\prime}(t)
\, . \label{awel}
\end{align}
Then the recursion equations can be converted into a single matrix
equation. This matrix equation has been solved by diagonalizing
some $2\times2$ matrix \cite{biham2000}. The eigenvalues of this
matrix are expressed as
\begin{equation}
\lambda_{\pm}=e^{\iu\omega_{\pm}}
\, , \qquad
\omega_{\pm}=\pi+\frac{\beta+\gamma}{2}\pm\omega
\, , \label{eigen0}
\end{equation}
where the parameter $\omega$ obeys $0\leq\omega\leq\pi$ and
\begin{equation}
\cos\omega=W_{k}\,\cos\frac{\beta+\gamma}{2}+W_{\ell}\,\cos\frac{\beta-\gamma}{2}
\ . \label{cosom}
\end{equation}
Under assumptions $W_{k}\neq0$ and $W_{\ell}\neq0$ together with
$\gamma\in(0;2\pi)$, the matrix of interest is certainly
diagonalizable. The averaged amplitudes are finally expressed as
\begin{align}
\wka^{\,\prime}(t)&=\xi_{1}e^{\iu\omega_{+}t}-\xi_{2}e^{\iu\omega_{-}t}
\, , \label{wkat}\\
\wel^{\,\prime}(t)&=\xi_{3}e^{\iu\omega_{+}t}-\xi_{4}e^{\iu\omega_{-}t}
\, , \label{welt}
\end{align}
where the coefficients are found as
$a=\bigl(1-e^{\iu\beta}\bigr)\,e^{\iu\gamma}W_{k}-e^{\iu\gamma}$,
$b=\bigl(1-e^{\iu\beta}\bigr)W_{\ell}$ and
\begin{align}
\xi_{1}&=\frac{(\lambda_{-}-a)\wka^{\,\prime}(0)-b\,\wel^{\,\prime}(0)}{\lambda_{-}-\lambda_{+}}
\ , &
\xi_{2}&=\frac{(\lambda_{+}-a)\wka^{\,\prime}(0)-b\,\wel^{\,\prime}(0)}{\lambda_{-}-\lambda_{+}}
\ , \nonumber\\
\xi_{3}&=\frac{\lambda_{+}-a}{b}\>\xi_{1}
\ , &
\xi_{4}&=\frac{\lambda_{-}-a}{b}\>\xi_{2}
\ . \label{cxi34}
\end{align}
Introducing the initial differences
\begin{align}
\Delta{k}_{x}^{\,\prime}&=k_{x}^{\,\prime}(0)-\wka^{\,\prime}(0)
\, , \label{delkdf}\\
\Delta\ell_{y}^{\,\prime}&=\ell_{y}^{\,\prime}(0)-\wel^{\,\prime}(0)
\, , \label{deledf}
\end{align}
the solution is completed by the formulas \cite{biham2000}
\begin{align}
k_{x}^{\,\prime}(t)
&=\wka^{\,\prime}(t)+(-1)^{t}e^{\iu\gamma{t}}\Delta{k}_{x}^{\,\prime}
\, , \label{delka1}\\
\ell_{y}^{\,\prime}(t)
&=\wel^{\,\prime}(t)+(-1)^{t}\Delta\ell_{y}^{\,\prime}
\, . \label{delel1}
\end{align}


\begin{thebibliography}{100}

\bibitem{grover97}
Grover, L.K.: Quantum mechanics helps in searching for a needle in a haystack. Phys. Rev. Lett. {\bf 79}, 325--328 (1997)

\bibitem{grover97a}
Grover, L.K.: Quantum computers can search arbitrarily large databases by a single query. Phys. Rev. Lett. {\bf 79}, 4709--4712 (1997)

\bibitem{grover98}
Grover, L.K.: Quantum computers can search rapidly by using almost any transformation. Phys. Rev. Lett. {\bf 80}, 4329--4332 (1998)

\bibitem{shor97}
Shor, P.W.:  Polynomial-time algorithms for prime factorization and discrete logarithms on a quantum computer. SIAM J. Comput. {\bf 26}, 1484--1509 (1997)

\bibitem{hama2006}
Haase, D., Maier, H.: Quantum algorithms for number fields. Fortschr. Phys. {\bf 54}, 866--881 (2006)

\bibitem{hallg07}
Hallgren, S.: Polynomial-time quantum algorithms for Pell's equation and the principal ideal problem. J. ACM {\bf 54}, 4 (2007)

\bibitem{vandam2010}
Childs, A.M., van Dam, W.: Quantum algorithms for algebraic problems. Rev. Mod. Phys. {\bf 82}, 1--52 (2010)

\bibitem{loka2007}
Lomonaco, S.J., Kauffman, L.H.: Is Grover's algorithm a quantum hidden subgroup algorithm? Quantum Inf. Process. {\bf 6}, 461--476 (2007)

\bibitem{bbbv97}
Bennett, C.H., Bernstein, E., Brassard, G., Vazirani, U.: Strengths and weaknesses of quantum computing. SIAM J. Comput. {\bf 26}, 1510--1523 (1997)

\bibitem{zalka99}
Zalka, C.: Grover's quantum searching algorithm is optimal. Phys. Rev. A {\bf 60}, 2746--2751 (1999)

\bibitem{patel2016}
Patel, A.D., Grover, L.K.: Quantum search. In: Kao, M.-Y. (ed.) Encyclopedia of Algorithms, pp. 1707--1716. Springer, New York (2016)

\bibitem{biham99}
Biham, E., Biham, O., Biron, D., Grassl, M., Lidar, D.A.: Grover's
quantum search algorithm for an arbitrary initial amplitude
distribution. Phys. Rev. A {\bf 60}, 2742--2745 (1999)

\bibitem{biham2000}
Biham, E., Biham, O., Biron, D., Grassl, M., Lidar, D.A., Shapira,
D.: Analysis of generalized Grover quantum search algorithms using
recursion equations. Phys. Rev. A {\bf 63}, 012310 (2000)

\bibitem{biham2002}
Biham, E., Kenigsberg, D.: Grover's quantum search algorithm for
an arbitrary initial mixed state. Phys. Rev. A {\bf 66}, 062301
(2002)

\bibitem{deutsch85}
Deutsch, D.: Quantum theory, the Church--Turing principle and the universal quantum computer. Proc. R. Soc. Lond. A {\bf 400}, 97--117 (1985)

\bibitem{bpati2002}
Braunstein, S.L., Pati, A.K.: Speed-up and entanglement in quantum searching. Quantum Inf. Comput. {\bf 2}, 399--409 (2002)

\bibitem{jozsa03}
Jozsa, R., Linden, N.: On the role of entanglement in quantum-computational speed-up. Proc. R. Soc. Lond. A {\bf 459}, 2011--2032 (2003)

\bibitem{bcp14}
Baumgratz, T., Cramer, M., Plenio, M.B.: Quantifying coherence. Phys. Rev. Lett. {\bf 113}, 140401 (2014)

\bibitem{abc16}
Adesso, G., Bromley, T.R., Cianciaruso, M.: Measures and applications of quantum correlations. J. Phys. A: Math. Theor. {\bf 49}, 473001 (2016)

\bibitem{plenio16}
Streltsov, A., Adesso, G., Plenio, M.B.: Quantum coherence as a resource. Rev. Mod. Phys. {\bf 89}, 041003 (2017)

\bibitem{fan2017}
Hu, M.-L., Hu, X., Peng, Y., Zhang, Y.-R., Fan, H.: Quantum coherence and quantum correlations. E-print arXiv:1703.01852 [quant-ph] (2017)

\bibitem{hillery16}
Hillery, M.: Coherence as a resource in decision problems: The Deutsch-Jozsa algorithm and a variation. Phys. Rev. A {\bf 93}, 012111 (2016)

\bibitem{hfan2016}
Shi, H.-L., Liu, S.-Y., Wang, X.-H., Yang, W.-L., Yang, Z.-Y.,
Fan, H.: Coherence depletion in the Grover quantum search
algorithm. Phys. Rev. A {\bf 95}, 032307 (2017)

\bibitem{apati2016}
Anand, N., Pati, A.K.: Coherence and entanglement monogamy in the discrete analogue of analog Grover search. E-print arXiv:1611.04542 [quant-ph] (2016)

\bibitem{farhi98}
Farhi, E., Gutmann, S.: Analog analogue of a digital quantum computation. Phys. Rev. A {\bf 57}, 2403--2406 (1998)

\bibitem{nielsen}
Nielsen, M.A., Chuang, I.L.: Quantum Computation and Quantum Information. Cambridge University Press, Cambridge (2000)

\bibitem{watrous1}
Watrous, J.: The Theory of Quantum Information. Cambridge University Press, Cambridge (2018) 

\bibitem{vedral02}
Vedral, V.: The role of relative entropy in quantum information theory. Rev. Mod. Phys. {\bf 74}, 197--234 (2002)

\bibitem{pati16}
Singh, U., Pati, A.K., Bera, M.N.: Uncertainty relations for quantum coherence. Mathematics {\bf 4}, 47 (2016)

\bibitem{pzflf16}
Peng, Y., Zhang, Y.-R., Fan, Z.-Y., Liu, S., Fan, H.: Complementary
relation of quantum coherence and quantum correlations in multiple
measurements. E-print arXiv:1608.07950 [quant-ph] (2016)

\bibitem{rastcomu}
Rastegin, A.E.: Uncertainty relations for quantum coherence with respect to mutually unbiased bases. Front. Phys. {\bf 13},  130304 (2018)

\bibitem{rastpra16}
Rastegin, A.E.: Quantum coherence quantifiers based on the Tsallis relative $\alpha$ entropies. Phys. Rev. A {\bf 93}, 032136 (2016)

\bibitem{shao16}
Shao, L.-H., Li, Y., Luo, Y., Xi, Z.: Quantum coherence quantifiers based on the R\'{e}nyi $\alpha$-relative entropy. Commun. Theor. Phys. {\bf 67}, 631--636 (2017)

\bibitem{skwgb16}
Streltsov, A., Kampermann, H., W\"{o}lk, S., Gessner, M., Bru\ss,
D.: Maximal coherence and the resource theory of purity. New J. Phys. {\bf 20}, 053058 (2018)

\bibitem{hall15}
Cheng, S., Hall, M.J.W.: Complementarity relations for quantum coherence. Phys. Rev. A {\bf 92}, 042101 (2015)

\bibitem{baietal6}
Yuan, X., Bai, G., Peng, T., Ma, X.: Quantum uncertainty relation using coherence. Phys. Rev. A {\bf 96}, 032313 (2017)

\bibitem{hufan15}
Hu, M.-L., Fan, H.: Evolution equation for quantum coherence. Sci. Rep. {\bf 6}, 29260 (2016)

\bibitem{shao15}
Shao, L.-H., Xi, Z., Fan, H., Li, Y.: Fidelity and trace-norm distances for quantifying coherence. Phys. Rev. A {\bf 91}, 042120 (2015)

\bibitem{rpl15}
Rana, S., Parashar, P., Lewenstein, M.: Trace-distance measure of coherence. Phys. Rev. A {\bf 93}, 012110 (2016)

\bibitem{uhlmann76}
Uhlmann, A: The 'transition probability' in the state space of a *-algebra. Rep. Math. Phys. {\bf 9}, 273--279 (1976)

\bibitem{jozsa94}
Jozsa, R.: Fidelity for mixed quantum states. J. Mod. Optics {\bf 41}, 2315--2323 (1994)

\bibitem{geomes17}
H.-J.~Zhang, B.~Chen, M.~Li, S.-M.~Fei and G.-L.~Long, Estimation on geometric measure of quantum coherence. Commun. Theor. Phys. {\bf 67}, 166-- (2017)

\bibitem{bera15}
Bera, M.N., Qureshi, T., Siddiqui, M.A., Pati, A.K.: Duality of quantum coherence and path distinguishability. Phys. Rev. A {\bf 92}, 012118 (2015)

\bibitem{bagan16}
Bagan, E., Bergou, J.A., Cottrell, S.S., Hillery, M.: Relations between coherence and path information. Phys. Rev. Lett. {\bf 116}, 160406 (2016)

\bibitem{qureshi17}
Qureshi, T., Siddiqui, M.A.: Wave-particle duality in $N$-path interference. Ann. Phys. {\bf 385}, 598--604 (2017)

\bibitem{hufan16}
Hu, M.-L., Fan, H.: Relative quantum coherence, incompatibility and quantum correlations of states. Phys. Rev. A {\bf 95}, 052106 (2017)

\bibitem{aasv2006}
Ambainis, A., Schulman, L.J., Vazirani, U.: Computing with highly mixed states. J. ACM {\bf 53}, 507--531 (2006)

\bibitem{popf2016}
Popescu, P., Slu\c{s}anschi, E.-I., Iancu, V., Pop, F.: A new
upper bound for Shannon entropy. A novel approach in modeling of
Big Data applications. Concurrency Computat.: Pract. Exper. {\bf
28}, 351--359 (2016)

\end{thebibliography}
\end{document}